\documentclass[%
 reprint,
 amsmath,amssymb,
 aps,
prb,
]{revtex4-2}
\usepackage{float}
\usepackage{graphicx}
\usepackage{dcolumn}
\usepackage{bm}
\usepackage{braket}
\usepackage{xcolor}
\makeatletter
\let\newfloat\newfloat@ltx
\makeatother
\usepackage{algorithm}
\usepackage{algpseudocode}
\usepackage[normalem]{ulem}
\usepackage[colorlinks=true]{hyperref}

\begin{document}
\title{Quantum Counting in the Rydberg Blockade}
\author{Joseph Gibson}
\affiliation{Institut quantique, Sherbrooke, Québec, J1K 2R1, Canada}
\affiliation{Département de physique, Université de Sherbrooke, Sherbrooke, Québec, J1K 2R1, Canada}
\author{Victor Drouin-Touchette}
\affiliation{Institut quantique, Sherbrooke, Québec, J1K 2R1, Canada}
\affiliation{Département de physique, Université de Sherbrooke, Sherbrooke, Québec, J1K 2R1, Canada}
\affiliation{Département de génie électrique et de génie informatique, Université de Sherbrooke, Sherbrooke, Québec, J1K 2R1, Canada}
\author{Stefanos Kourtis}
\affiliation{Institut quantique, Sherbrooke, Québec, J1K 2R1, Canada}
\affiliation{Département de physique, Université de Sherbrooke, Sherbrooke, Québec, J1K 2R1, Canada}
\affiliation{Département d'informatique, Université de Sherbrooke, Sherbrooke, Québec, J1K 2R1, Canada}
\affiliation{Mila -- Quebec AI Institute, Montréal, Québec, H2S 3H1, Canada}
\date{\today}

\begin{abstract}
We propose a quantum algorithm for approximately counting the number of solutions to planar 2-satisfiability (2SAT) formulas natively on neutral-atom quantum computers (NAQC).
Our algorithm maps Boolean variables to atomic registers arranged in space according to a given formula, so that 2SAT constraints are enforced via the Rydberg blockade between neighboring atoms.
A quench of an initial computational basis state then produces a superposition of all solutions after sufficiently long evolution.
For almost uniform superpositions, a polynomial number of measurements is enough to estimate the solution count up to any constant multiplicative factor via sampling-based counting.
Numerical simulations show that the protocol leads to almost uniform solution sampling in 1D and 2D grids and produces accurate counts for small random 2SAT instances on punctured grids. We implement our algorithm on Pasqal's Orion Alpha NAQC and demonstrate approximate counts within small multiplicative factors of the exact count for a planar grid instance with 81 atoms.
\end{abstract}

\maketitle
\section{Introduction}
Counting how many objects in a set have a certain property is a fundamental computational task~\cite{Valiant1979}. Counting problems lie at the heart of many interesting applications in artificial intelligence~\cite{Roth1996, sangPerformingBayesianInference2005, Baluta2019}, network reliability~\cite{Valiant1979, DuenasOsorio2017}, and statistical physics~\cite{ising_part, Jerrum1993}, to name a few. 
However, many counting problems of practical importance are intractable.
The computational complexity class \#P that encompasses counting contains problems that are thought to be much harder than the hardest problems in the more commonly encountered class NP~\cite{sharp_permanent, Valiant1979}.
An efficient exact algorithm for hard problems in \#P could be bootstrapped to solve any problem in the polynomial hierarchy efficiently~\cite{Todathm}.
The existence of such algorithms is hence considered unlikely. For this reason, counting algorithms whose runtime is exponential in the number of variables in the worst case but which perform reasonably well in practice are sought-after~\cite{gomes2021model}.

Exact counting methods are based on component caching~\cite{componentcaching}, knowledge compilation~\cite{knowledgec}, and tensor networks~\cite{stefanos_fast_count, dudekEfficientContractionLarge2020}.
While these algorithms greatly outperform naive elimination of non-solutions, they are limited to instances with up to a few hundred variables.
Approximate counting, in which we seek only to obtain the solution count up to a multiplicative factor, relies on either hashing or sampling the solution set.
Hashing-based counting involves randomly bisecting the solution set recursively using a hash function family and asking an NP oracle (e.g., a Boolean satisfiability solver) if the hashed set is empty or not~\cite{onapproximatingsharpP}.
Sampling-based techniques require that solutions be sampled at random almost uniformly~\cite{JVV_alg}.
This requirement is rather stringent, and hence hashing-based techniques have been more widely adopted to date~\cite{gomes2021model}.

It is natural to ask whether quantum computing can offer improvements in solving counting problems.
Early work~\cite{Gilles} based on the Grover iteration~\cite{Grover1996} and recent advances~\cite{Aaronson2020, aaronson2020quantum2, belovs2020tight, kretschmer2019mathsfqma} do provide an improvement in the black-box setting, but the circuit depths required to implement these algorithms grow exponentially in the number of variables, thus rendering them inaccessible to noisy quantum processors before fault tolerance~\cite{Stoudenmire2024_no_grover, lee2024implementation}.
Recently proposed hybrid quantum-classical algorithms for counting~\cite{Drapeau_2026, zhangGroverQAOA3SATQuadratic2024, sundarQuantumAlgorithmCount2019}, while more readily implementable in the near term, depend on variational search that is not guaranteed to converge~\cite{Bittel2021, larocca2024reviewbarrenplateausvariational}.

Here we introduce a quantum algorithm for counting solutions to planar 2-satisfiability (2SAT) formulas on existing neutral-atom quantum computers (NAQCs). 
In our algorithm, Boolean variables are represented by atomic registers and 2SAT constraints are enforced by the Rydberg interaction between neighboring atoms.
Each iteration consists of (i) initial placement of atoms according to a given 2SAT formula, (ii) quench with a static laser pulse, and (iii) measurement in the computational basis.
If we also assume that arbitrary initial computational basis states can be prepared, the measured samples can be used as initial states, leading to a feed-forward (FF) protocol.
We show numerically that our algorithm requires $O(\mathrm{poly}(n) \log(n))$ operations to perform sampling-based counting on instances with $n$ variables, returning upper and lower bounds on the exact count for punctured 2D graphs.
Finally, we implement our algorithm on Pasqal's Orion Alpha quantum processing unit (QPU) and tackle a planar \#2SAT instance with 81 variables defined on a triangular grid. Even without the FF protocol, we obtain approximate counts within a small multiplicative factor of the exact count.

\section{Theory}

\subsection{Model for neutral-atom hardware}

A NAQC consists of alkali atoms trapped in optical tweezers and positioned so as to represent a computational problem. 
Atoms are then cooled to their electronic ground state.
For example, we can consider Rb atoms, whose ground state is $|0\rangle = |5S_{1/2}\rangle$.
A laser then coherently drives atoms to the Rydberg state $|1\rangle = |60S_{1/2}\rangle$.
Strong dipole-dipole interactions between nearby atoms in the Rydberg state create entanglement and power analog quantum computation and simulation~\cite{wintersperger2023neutral, Browaeys2020}.

The dynamics of a register of $n$ trapped atoms is described as evolution under the Rydberg Hamiltonian~\footnote{Here we assume that the driving laser is tuned exactly to the resonance between ground and Rydberg states. We therefore omit the detuning term that commonly appears in the Rydberg Hamiltonian.}

\begin{equation}
    H_{\text{Ryd}}(\Omega) = \sum_{i=1}^n \frac{\Omega}{2}X_i + \sum_{i<j}^n V_{ij} n_i n_j \; ,
    \label{eq:rydberg_ham}
\end{equation}
where $X_i = |1_i\rangle\langle 0_i| + |0_i\rangle\langle 1_i|$ describes Rabi oscillations between ground (excited) state $\ket{0_i}$ ($\ket{1_i}$) for atom $i$ at the resonance frequency $\Omega$, $V_{ij} \approx C_6 r_{ij}^{-6}$ is the interaction between atoms $i$ and $j$ in the Rydberg state at distance $r_{ij}$ with $C_6>0$, and $n_i = |1_i\rangle\langle 1_i|$.

In the limit $V_{ij} \gg \Omega$, called the Rydberg blockade, pairs of atoms within a distance smaller than the blockade radius $r_B = (C_6/\Omega)^{1/6}$ are energetically prohibited from simultaneously occupying the Rydberg state $\ket{1}$.
Given the sharp decay of interactions with distance, the atomic array effectively forms a so-called unit-disk graph $G = (V,E)$, where the $i$-th atom is represented by vertex $i \in V$ ($|V| = n$) and pairs of atoms $i$ and $j$ with $r_{ij} < r_B$ are connected by edges $(i,j) \in E$ of equal length.

\subsection{Encoding 2SAT in the Rydberg blockade}

Without nearest-neighbor excitations, the low-energy states of the Rydberg blockade directly correspond to the solution space of a family of 2SAT formulas.
If we denote the computational basis states of two neighboring atoms within a blockade radius as $\ket{x_i x_j}$, where $x_i, x_j \in \{0,1\}$ are Boolean variables, we see that low-energy states satisfy the constraint $\neg x_i \lor \neg x_j$, where $\neg$ is the negation operation $\neg x_i = 1-x_i$.
Therefore, every component $\ket{x} = \ket{x_n \dots x_1}$ of a low-energy state in the Rydberg blockade is a solution to a monotone 2SAT formula
\begin{equation}
\label{eq:satcnf}
    \phi_G(x) = \bigwedge_{(i,j)\in E} (\neg x_i \lor \neg x_j) \; ,
\end{equation}
defined for a given unit-disk graph $G$. Computing the number of solutions $\# \phi_G = |\{ x : \phi_G(x)=1 \}|$ of arbitrary monotone 2SAT formulas is \#P-complete, i.e., as hard as the hardest problems in the class \#P~\cite{Valiant1979}, and remains so even for planar 3-regular bipartite graphs~\cite{xia2006}.

\subsection{Sampling-based counting}

To design a quantum algorithm for counting, we exploit a well-known correspondence between almost uniform sampling and approximate counting, first operationalized in a classical algorithm by Jerrum, Valiant, and Vazirani (JVV)~\cite{JVV_alg}.
The JVV algorithm samples solutions to a given problem, and then fixes a variable to its observed maximum-likelihood state in the collected samples.
The variable fixing defines the subproblem whose solutions we sample in the next step.
This process is called self-reduction and continues until all variables have been fixed~\footnote{A visual representation of the self-reduction process can be found in Ref.~\cite{moore_computation}.}. 
An estimate of the solution count is computed from the product of the reciprocals of all observed maximum likelihoods.
There are two requirements for the estimate to be within a constant multiplicative factor of the exact count: the problem must be self-reducible, and solutions must be sampled sufficiently uniformly.

A Boolean formula $\phi$ is self-reducible if a solution can be constructed out of solutions to subformulas of the problem, where the subformulas are defined by fixing Boolean variables. This can be expressed as
\begin{equation}
    \phi(x) = 1 \Leftrightarrow \{\phi(x | x_i = 1)=1\} \vee \{\phi(x | x_i = 0) = 1 \} \; .
\end{equation}
All SAT problems are self-reducible~\cite{self_red_def}, and the first requirement is hence satisfied in our case.

To assess how uniform a solution sampler is, we can compare its output distribution $D$ with the uniform distribution $U$ over the solution set $\mathcal{X} = \{x |\phi(x) = 1\}$. One way to quantify the sampler non-uniformity is via the total variation distance~\cite{moore_computation}
\begin{equation}
    \eta (D) = ||D-U||_{\mathrm{TV}} = \frac{1}{2}\sum_{x\in \mathcal{X}} |D(x)-U(x)| \; . \label{eq:nonunif}
\end{equation}
A sampler that satisfies $n\eta = O(1)$ can be used to estimate the count up to an arbitrarily small multiplicative factor for any instance of size $n$~\cite{JVV_alg}.

\subsection{Counting using the Rydberg blockade}

We now ask whether these requirements can be satisfied in NAQCs. First, we show that this representation of 2SAT problems preserves self-reducibility. Suppose that in the self-reduction process we wish to fix the $c$-th variable to $x_c = 0$. On a NAQC, we must fix $|x_c \rangle = |0\rangle$. According to~\eqref{eq:satcnf}, an atom in its ground state does not constrain the state of its first nearest neighbors in $G=(V,E)$. We can therefore equivalently represent the reduced instance with an array that simply omits atom $c$ altogether, that is, the induced subgraph $G'=(V',E')$ with $V'=V \setminus \{c\}$ and $E'$ adjusted accordingly. Now suppose that we fix $x_c = 1$. This choice forces the \textit{first} neighbors of $c$ to be in $\ket{0}$, which in turn means that the \textit{second} neighbors of $c$ are unconstrained by the choice $|x_c \rangle = |1\rangle$. We can therefore represent the reduced instance with an array that omits atom $c$ and its first nearest neighbors, i.e., $G''=(V'',E'')$ with $V'' = V \setminus (\{c\} \, \cup \, \{ j | (c,j) \in E \})$ and $E''$ adjusted accordingly.

The next question is how uniformly we can sample solutions using a NAQC. This question is open-ended, since one can devise many protocols to explore the low-energy space of~\eqref{eq:rydberg_ham}.
For all but an $O(1)$ subset of initial states that exhibit unexpectedly long thermalization times~\cite{Probing_51_atom_qs, Scared_eigenstates_turner, bluvstein2021controlling}, the dynamics induced by evolution under~\eqref{eq:rydberg_ham} is thought to be chaotic~\cite{choi2023preparing, Ho_2019chaos, Scared_eigenstates_turner, schnee2024unconventional}.
In the limit $t \rightarrow \infty$, generic initial states spread maximally in the accessible Hilbert space. In practice, evolution times will be upper-bounded by qubit coherence.
A quench of an initial computational basis state in the Rydberg blockade will not produce a uniform distribution over the low-energy subspace at the short times accessible on today's QPUs. Approximate uniform sampling of the solution space of \eqref{eq:satcnf} thus requires a protocol adapted to the limitations of existing hardware.

\section{Algorithm}

We now introduce \textsc{RydCount}, a sampling-based counting algorithm for NAQCs, as described in Algorithm~\ref{alg:RQC}. \textsc{RydCount} takes as input the register \(R\) that defines a 2SAT instance in its blockade graph, the number of samples \(n_{\text{samp}}\), the Hamiltonian parameters (Rabi frequency \(\Omega\)), evolution time bounds \(t_{\mathrm{min}}, t_{\mathrm{max}}\), and an integer $k$ to be defined below.
The core of \textsc{RydCount} is repeated iterations of a subroutine we call \textsc{RydSamp} to gather samples (steps 4--13). The algorithm then follows the JVV framework and calculates the variable \(x_c\) that is most likely to be 1 (steps 14 and 15). 
The variable \(x_c\) is fixed to 1, the corresponding atom (step 16) and all its nearest neighbors (step 17) are removed from the array, and the process repeats on the reduced array.
Finally, the estimated count \(\lambda\) is updated (step 18) until all variables are fixed. 

\begin{algorithm}[t]
\caption{\textsc{RydCount}}
\label{alg:RQC}
\textbf{Input: }\(R\): atom register expressing \(\phi_{G}(x)\) as the blockade graph \(G=(V,E)\)
\\ \(\Omega\): Rabi frequency 
\\ \(n_{\mathrm{samp}}\): number of samples taken at each step
\\ \(k\): number of feed-forward steps
\\ \(t_{\mathrm{min}}, t_{\mathrm{max}}\): minimum and maximum evolution times
\begin{algorithmic}[1]
\State \(\lambda \gets 1\) 
\Repeat
\State \(\sigma \gets \emptyset\)
\State \(\ket{\psi_0} \gets \ket{0}^{\otimes |V|}\)
\For{\(m = 1 \dots n_\mathrm{samp}\)} \Comment{This is \textsc{RydSamp}}
\For{\(j = 1 \dots k\)} \Comment{Feed-Forward loop}
\State \(t_{\mathrm{rand}} \xleftarrow{\text{sample}} U(t_{\mathrm{min}}, t_{\mathrm{max}})\) \Comment{\(U\): uniform dist.}
\State Initialize register \(R\) in state \(\ket{\psi_0}\)
\State \(x \xleftarrow{\text{measure}} \exp(-iH_{\mathrm{Ryd}}(\Omega)t_{\mathrm{rand}})\ket{\psi_0}\) 
\State \(\ket{\psi_0} \gets \ket{x}\) \Comment{If: \(\phi_G(x)\)}
\EndFor
\State \(\sigma \gets \sigma \cup \{x\}\)
\EndFor
\State \(p_i \gets \frac{1}{n_{\mathrm{samp}}} \sum_{x\in \sigma} x_i\) \Comment{likelihood of \(x_i\)}
\State \(c \gets \operatorname{argmax}_i(p_i)\) 
\State \(R \gets R(x \mid x_c = 1)\) \Comment{Reduce the register}
\State \(R \gets R(x \mid x_j = 0 \; \forall \; (c,j) \in E)\) \Comment{Reduce the register}
\State \(\lambda \gets \frac{\lambda}{p_c}\)
\Until{$x_i \gets \{0,1\} \: \forall i \in V$} 
\end{algorithmic}
\textbf{Output:} Estimated count \(= \lambda\)
\end{algorithm}

The key part of this algorithm is \textsc{RydSamp}. This is the preparation and measurement of the system after an evolution $\mathcal{U} = \exp[-iH_{\mathrm{Ryd}}(\Omega)t]$ at time $t=t_{\mathrm{rand}}$ taken uniformly at random in the time interval $[t_{\mathrm{min}}, t_{\mathrm{max}}]$. After one iteration of \textsc{RydSamp}, if the measured bitstring is a valid solution to $\phi_G$, that bitstring is fed in as the new initial state (step 10 in Algorithm~\ref{alg:RQC}). The $k$-th step of this \textit{feed-forward} (FF) protocol is the channel
\begin{align}
    D_{k}(x) &= \text{Tr}(P_x \mathcal{U}\rho_{k-1} 
    \mathcal{U}^\dagger), \\
    \rho_{k-1} &= \sum_{x \in \mathcal{X}} D_{k-1}(x) \ket{x}\bra{x},
\end{align}
where $\rho_k$ is the density operator at step $k$, $\rho_0 = \ket{0}^{\otimes n} \bra{0}^{\otimes n}$, \(P_x = \ket{x}\bra{x}\) is the projection operator onto state $x \in \mathcal{X}$, where $\mathcal{X}$ is the set of solutions of $\phi_G$. If the evolution under $\mathcal{U}$ is mixing,  $\rho_k \rightarrow \frac{\mathbb{I}}{|\mathcal{X}|}$ as $k \rightarrow \infty$ and the sampling is asymptotically uniform.

As a concrete example, we characterize \textsc{RydSamp} by classically simulating its application to a 1D instance. We set $\Omega =1$, $V_{ij}=50$ for nearest neighbors, with an $r^{-6}$ decay for all further neighbors~\footnote{All calculations were performed using the QuSpin library~\cite{weinberg_quspin} on an AMD EPYC 7F72 CPU with 100GB of RAM available.}.
The minimum evolution time is chosen to be $t_{\mathrm{min}} \approx \frac{10}{\Omega}$. This choice ensures that the evolution is sufficiently long for the initial state amplitude to spread in the solution space (see  Appendix~\ref{appendix:min_time} for additional details).
The maximum evolution time is empirically chosen to be realistic for current physical devices, $t_{\mathrm{max}} \approx \frac{100}{\Omega}$~\cite{senoo2025_highfidelityentanglementcoherentmultiqubit, PasqalOrion,Wintersperger_2023naqccoherence}.

\begin{figure}
    \centering
    \includegraphics[width=\linewidth]{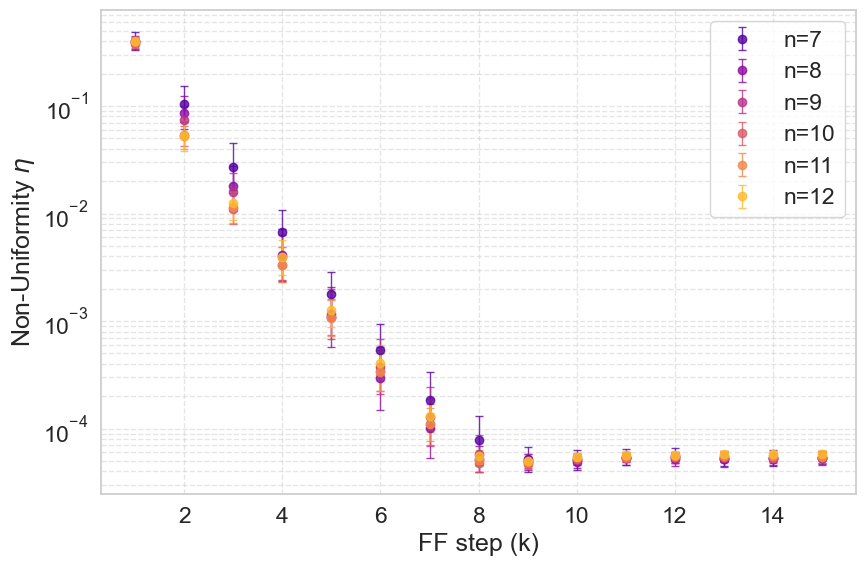}
    \caption{Nonuniformity $\eta$ as a function of number of FF steps for \textsc{RydSamp} applied to a 1D chain instance.}
    \label{fig:tvd_vs_step}
\end{figure}

Figure~\ref{fig:tvd_vs_step} shows an exponential decay in the nonuniformity as a function of the FF step, albeit bounded by a minimum plateau. This plateau is caused by two factors. The first is leakage out of the low-energy solution subspace due to the finiteness of the nearest-neighbor interaction that renders the Rydberg blockade imperfect. This can be mitigated, to some extent, by reducing the interatomic distance and thus strengthening the Rydberg blockade. The second is an energetic bias due to the long-range tail of the dipole-dipole interaction that penalizes high Hamming weight states. 
If we treat the channel $D_{k}(x)$ as a Markov chain, the transition matrix on the restricted subspace of solution states is not doubly stochastic and thus the desired convergence to a uniform distribution is not possible under the Rydberg Hamiltonian due to these factors.
This means that \textsc{RydSamp} converges to a stationary distribution that is only approximately uniform, with an \(\mathcal{O}(\log n)\) overhead in the number of FF steps.
Crucially, however, for the parameters chosen here, \textsc{RydSamp} achieves almost uniform sampling with $n\eta \lesssim 1$ for \#2SAT instances with $n\lesssim 10^4$, an order of magnitude larger than the largest instances typically accessible to classical solvers today (see Appendix~\ref{appendix:classical}).

In contrast, in the limit of infinite-strength, nearest neighbor-only interactions, i.e., the so-called PXP limit, there is no nonuniformity plateau (see Appendix~\ref{appendix:PXP}).

\section{Results}

\begin{figure*}[]
    \centering
    \includegraphics[width=\textwidth]{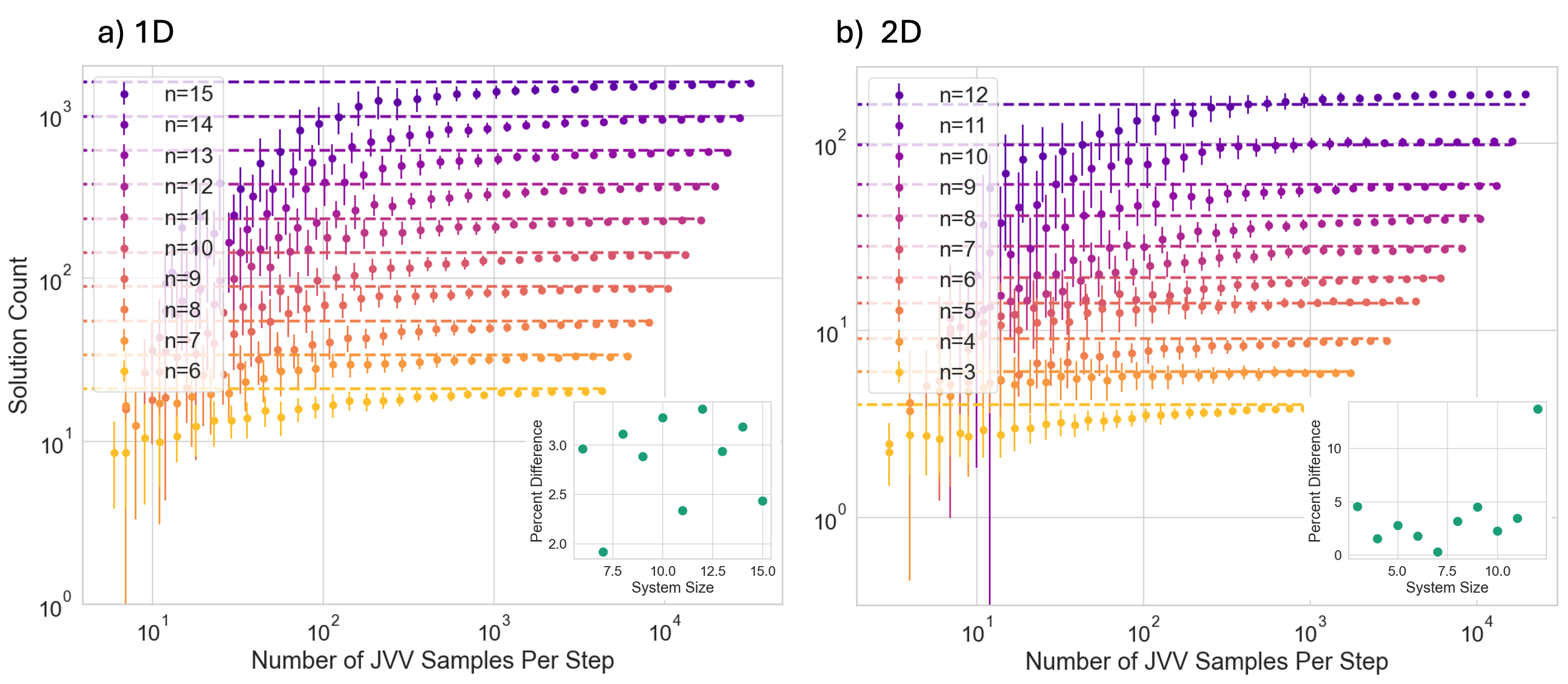}
    \caption{Accuracy of \textsc{RydCount} as a function of $n_{\text{samp}}$ with 15 FF steps evolving under the Rydberg Hamiltonian for (a) 1D chains of length $n$ and (b) 2D triangular grids of size $n$. Dashed lines represent exact counts. Insets show the relative error between estimated and exact counts versus system size.}
    \label{fig:1D_square_results}
\end{figure*}

\subsection{Classical simulation: 1D and 2D grids}

Next, we perform end-to-end benchmarks of \textsc{RydCount}.
We compare the estimated count to the exact count as a function of the number of samples $n_{\text{samp}}$ taken at each self-reduction step. For a sampler that satisfies $n\eta = \mathcal{O}(1)$, the theoretical worst-case bound for the convergence of the estimated count is $\mathcal{O}(\frac{n^2}{\epsilon^2 } \log(\frac{1}{\delta}))$, with relative error $\epsilon$ and failure probability $\delta$~\cite{JVV_alg, moore_computation, Samptocount}. 
Accordingly, our \textsc{RydCount} implementations for 1D and 2D grids configure the per-step number of samples via this bound, fixing $\epsilon = 0.2$ and $\delta = 0.05$.

In 1D, the solution count is known analytically~\cite{Feiguin_2007}.
Results with $k=15$ are shown in Fig.~\ref{fig:1D_square_results}(a). 
We observe that the estimate converges once $n_{\text{samp}}$ reaches $\frac{n^2}{\epsilon^2 } \log(\frac{1}{\delta}) = 35.52 n^2$ and that the error is well within the $1+\epsilon = 1.2$ multiplicative factor for all system sizes.

We then apply \textsc{RydCount} to 2D triangular grids and study its convergence to the exact count obtained with an exact \#SAT solver~\cite{imms-sat18} in Fig.~\ref{fig:1D_square_results}(b). We observe increasing error in the estimated count with system size. This is explained by the biasing factors described in the preceding section, namely, next-nearest neighbor interactions bias the sampling distribution away from states with a large Hamming weight. 
Thus, nodes measured in the $\ket{1}$ state will be less frequent, leading to an underestimate of the variable $p_i$ (step 14 in Algorithm~\ref{alg:RQC}).
The estimated count is iteratively found through division of $p_i$ after each step and thus leads to a systematic overestimation of the expected number of solutions. 
Therefore, if we choose to always fix variables to 1 in the self-reduction, \textsc{RydCount} will output an upper bound on the exact count. This is consistent with previous work using a biased sampler to bound rather than estimate within a multiplicative factor the exact count~\cite{Samptocount}. As in 1D, in the idealized case of a perfect Rydberg blockade, the bias disappears and \textsc{RydCount} provides an accurate estimate of the count (see Appendix~\ref{appendix:PXP}).

\subsection{Classical simulation: punctured 2D grids}

It is known that classical solvers cannot efficiently solve \#2SAT on graphs of maximum degree $\Delta \geq 6$ unless RP=NP~\cite{jerrum2002_deg6, sly2010computationaltransitionuniquenessthreshold}.
This class includes 2D triangular and 3D cubic grids, which are both experimentally accessible on NAQCs.
Here we will focus on non-trivial \#2SAT instances on planar triangular grids with a random selection of sites removed, i.e., punctured grids. While the exact complexity of \#2SAT for this ensemble is unknown to us, we tested the state-of-the-art classical counter GANAK~\cite{ganak2019} on random instances and found that the median runtime for low hole densities scaled exponentially with instance size (see Appendix~\ref{appendix:classical}).
This suggests that there are classically hard \#2SAT instances within this class of graphs.

\begin{figure}[b]
    \centering
    \includegraphics[width=0.9\linewidth]{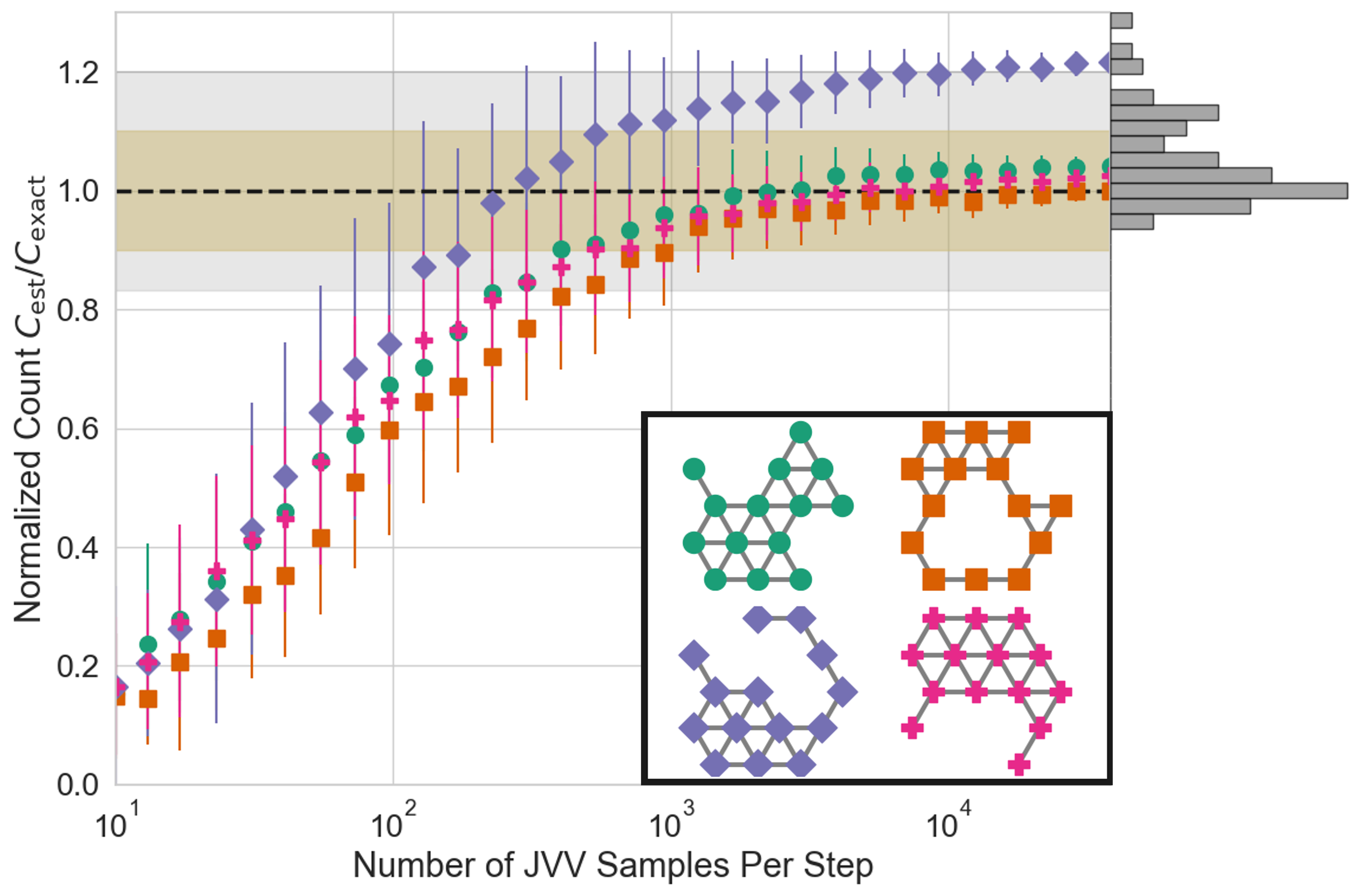}
    \caption{Ratio between the \textsc{RydCount} estimate $C_{\text{est}}$ and the exact count $C_{\rm exact}$ versus number of samples per self-reduction step for the 4 random punctured triangular grid instances shown in the inset with matching colors and markers. Shaded areas surrounding $C_{\text{est}}/C_{\text{exact}}=1$ correspond to approximate counts with a multiplicative factor of $1.1$ (light yellow) and $1.2$ (light grey). To obtain the histogram, we generate random instances where $4$ vertices are uniformly randomly chosen and removed from a $n=18$ grid. \textsc{RydCount} is performed on 100 of these random instances.}
    \label{fig:2d_jvv_results_punched}
\end{figure}

Fig.~\ref{fig:2d_jvv_results_punched} shows the convergence of the \textsc{RydCount} algorithm towards the exact count for triangular punctured graphs with low hole densities. For most graphs the algorithm converges to within a factor of 1.1 of the exact solution. However, the distribution of final counts (right histogram) exhibits a systematic bias, with most instances overestimating $C_{\text{exact}}$. The bias is a result of the next nearest neighbor interactions as before and is expected to become more pronounced as the system size grows. Consequently, for larger instances, \textsc{RydCount} acts as an effective upper bound on the exact count. Once again, in the idealized case of perfect Rydberg blockade, the bias disappears (see Appendix~\ref{appendix:PXP}).

\subsection{Experiment: \#2SAT instance on Pasqal QPU}

We cannot perform the full \textsc{RydCount} algorithm as originally formulated on existing analog NAQCs, as it requires selective atom addressability or hyperfine states, capabilities available in the lab but not on commercial devices~\cite{Bluvstein2025_localcontrol}.
We have nevertheless succeeded in implementing a simplified variant of \textsc{RydCount} with three modifications.
First, lacking selective atom addressability, we can only implement a single-step ($k=1$) version of \textsc{RydCount} starting from the ground state $\ket{0}^{\otimes n}$. In this case the resulting sampling distribution is not expected to satisfy $n \eta = \mathcal{O}(1)$.
Second, instead of dynamically selecting the optimal variable $x_c$ from measurement samples and performing $n$ self-reduction steps to fully solve an instance, we preselect a self-reduction order along a graph separator that decomposes the instance into two smaller subproblems. While the shortest separator in grids is obvious, short separators can also be found efficiently in arbitrary graphs~\cite{lipton1979separator}.
These subproblems are rapidly solved using a classical solver and the overall solution count is obtained by combining the subproblem counts with the quantum-derived multiplicative factor from the separator.
Third, to account for systematic overcounting in our algorithm, we evaluated two decimation paths. One path computes estimates and decimates along the preselected separator taking the $x_c = 1$ branch for all fixed variables, while the other takes the $x_c = 0$ path (one can easily adapt Algorithm~\ref{alg:RQC} to this path). Due to the sampling bias away from high Hamming-weight states, these two paths yield upper and lower bounds for the estimate. This protocol was deployed on a Pasqal Orion Alpha QPU located in Sherbrooke, Quebec, and used to solve the 81-variable \#2SAT instance shown in Fig.~\ref{fig:hardware_prob} (see Appendix~\ref{appendix:exp} for more details on the experiment). The results are shown in Fig.~\ref{fig:hardware_count}.

\begin{figure}
    \centering
    \includegraphics[width=\linewidth]{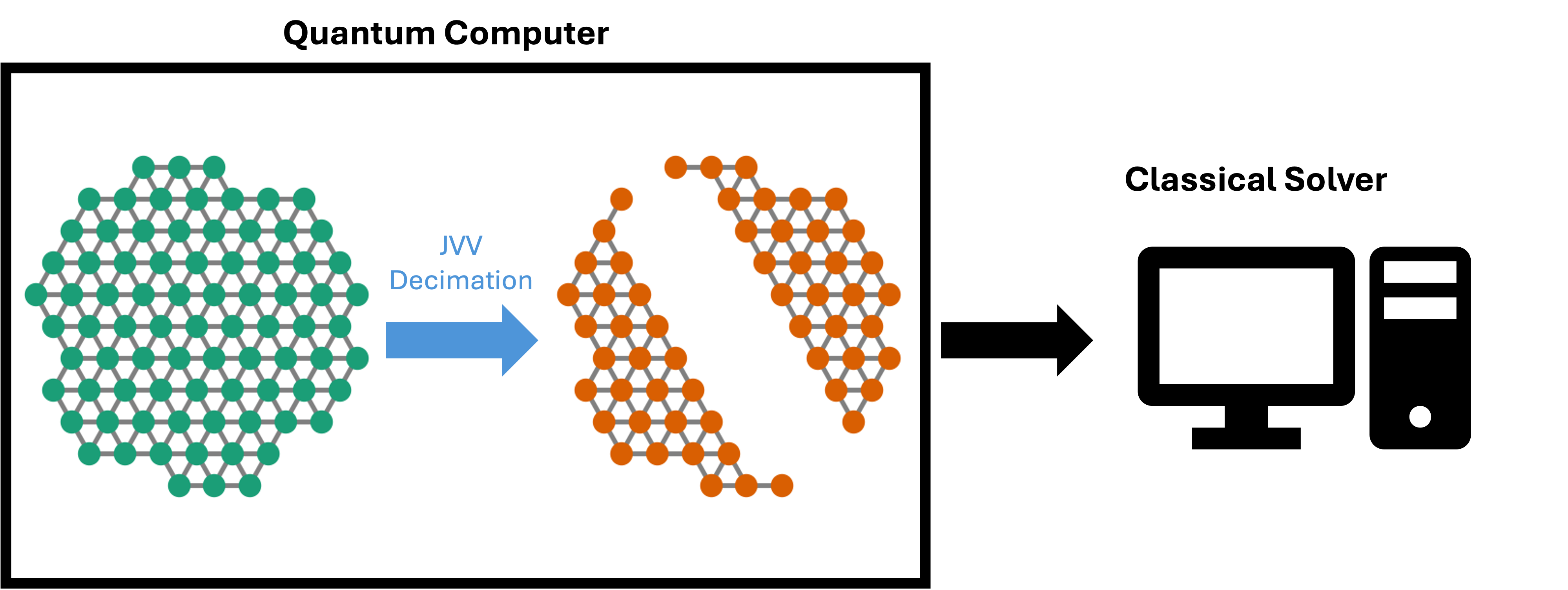}
    \caption{Sketch of the hybrid \textsc{RydCount} variant, in which self-reduction is performed on a QPU until the instance is split into two subproblems to be solved on a classical computer.}
    \label{fig:hardware_prob}
\end{figure}

To evaluate performance on hardware, we analyze both the raw shots from the QPU and a postselected dataset restricted solely to valid solution bitstrings (averaging $41\%$ sample retention per step). 
For the raw shots, two sets of points correspond to self-reduction paths where all variables are fixed to 1 (red hue) or to 0 (blue hue). For both paths, the count converges to an estimate at $\sim 10^3$ samples per JVV step. The two procedures result in counts that are within a multiplicative factor of 5 and 1.01, respectively, of the exact count.

Postselecting only valid independent sets to compute counts leads to worse estimates, indicating that the excellent estimate for the fix-to-0 path is likely a coincidence.
Nevertheless, even the postselected fix-to-0 path yields an estimate within a factor of less than $1.25$ of the exact count. Moreover, the correct count is bounded from above and below by the two postselected estimates, the $x_c = 0$ and $x_c = 1$ decimation paths, thus yielding useful bounds even in this rudimentary experiment.


These results illustrate that even the simplest implementation of \textsc{RydCount} on commercially available NAQCs already recovers a reasonably precise count for a complex computational problem with a number of variables ($n=81$) that may be considered ``large'' by the standards of current non-annealing QPUs.

\begin{figure}
    \centering
    \includegraphics[width=1\linewidth]{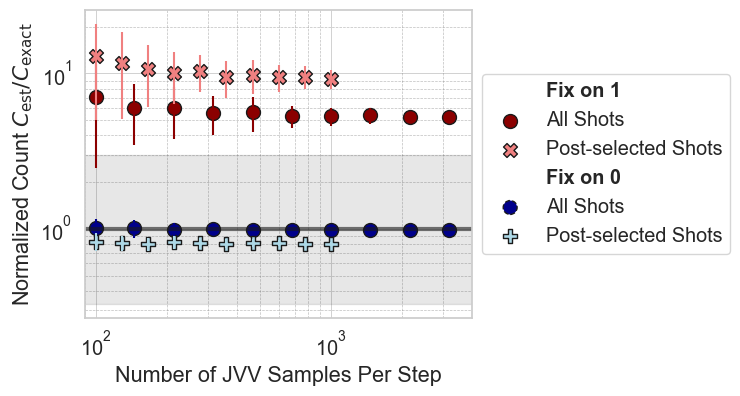}
    \caption{Convergence of hybrid \textsc{RydCount} variant for the 81-variable \#2SAT instance shown in Fig.~\ref{fig:hardware_prob}. Red (blue) hues correspond to the $x_c =1$ ($x_c = 0$) self-reduction path. The $x$-axis represents the number of samples per self-reduction step used for the estimate among the total set of samples collected from the hardware. The $y$-axis is the relative solution count $C_{\rm est}/C_{\rm exact}$. Shaded region corresponds to multiplicative error within a factor of $3$. Circles represent the count obtained using raw shots, while crosses represent the count obtained from postselected solution-only shots.}
    \label{fig:hardware_count}
\end{figure}

\section{Discussion and Outlook}

In summary, we have proposed an approximate counting algorithm that works natively on NAQCs.
Several near-term experimental advancements position \textsc{RydCount} as a competitor to state-of-the-art classical methods for challenging problems in \#P.
First, although the clock rate of NAQCs is currently less than $10$ Hz~\cite{PasqalOrion, quera_device}, rates of $100$ Hz or higher are achievable~\cite{dalyac2024graph} leading to faster overall sampling. 
Second, current NAQCs have hundreds of qubits~\cite{PasqalOrion, quera_device}, whereas \#SAT instances start challenging state-of-the-art classical solvers at several hundred to a few thousand variables. Trapping and rearrangement of registers of thousands of atoms has already been demonstrated~\cite{manetsch2024tweezerarray6100highly, pichard2024rearrangement,Schlosser2023_10000}. 
Third, \textsc{RydCount} and mitigation strategies for unwanted long-range interaction tails require selective addressability of atoms, which is already implemented in academic NAQCs~\cite{xiang2024observation, de2025demonstration, Browaeys2024_address}.
Finally, using chains of atoms that act as long edges, which have been demonstrated experimentally~\cite{Kim2022_wires}, we can encode \#P-complete instances as planar 3-regular bipartite graphs~\cite{xia2006}.

\begin{acknowledgments}
Parts of this research were financially supported by the National Research Council (NRC) Canada through its Applied Quantum Computing challenge program, Prompt through its \textit{Soutien aux organismes de recherche} program, the Natural Sciences and Engineering Research Council of Canada (NSERC) through an Alliance grant, and Pasqal Canada through its contribution to the Institutional Research Chair in Quantum Artificial Intelligence at Université de Sherbrooke. We thank the entire Pasqal Canada team, and in particular Mathieu Duval and Soufiane Kaghad, for their help with the hardware implementation.
\end{acknowledgments}

\appendix
\section{Choosing the minimum evolution time}
\label{appendix:min_time}

\begin{figure}[t]
    \centering
    \includegraphics[width=\linewidth]{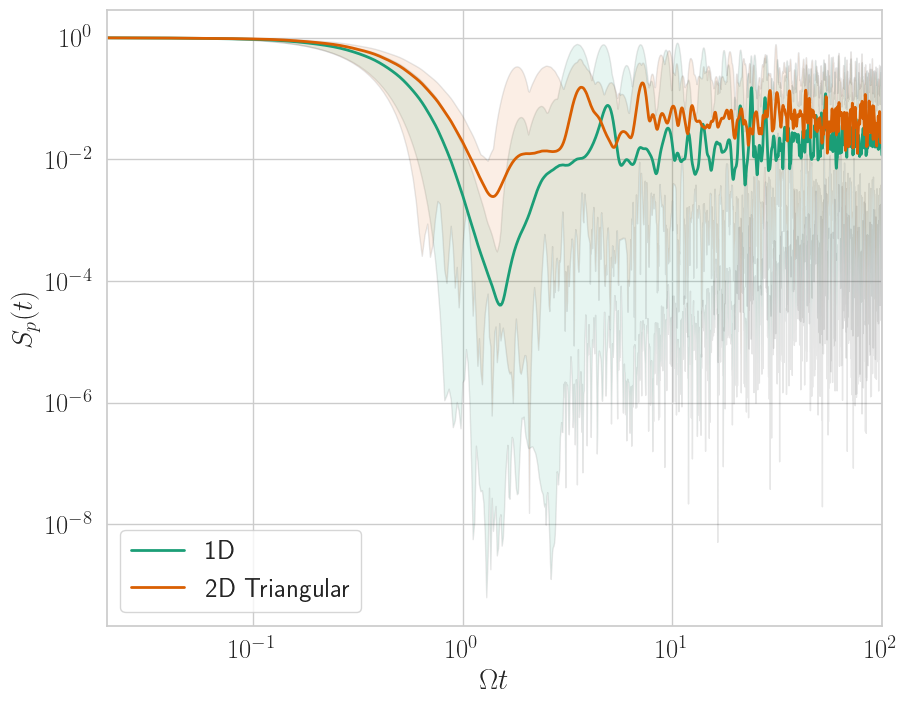}
    \caption{The average survival probability \(S_p(t) = |\bra{x}\mathcal{U}(t)\ket{x}|^2\) of all states in the solution set, with unitary evolution \(\mathcal{U}(t) = \exp(-itH_{\mathrm{Ryd}})\), for \(n=10\) on a 1D chain and a 2D triangular grid. Shaded regions show the range of the maximum and minimum values at each point in time for all of the evolutions.}
    \label{fig:initial_state}
\end{figure}

\begin{figure}[t]
    \centering
    \includegraphics[width=\linewidth]{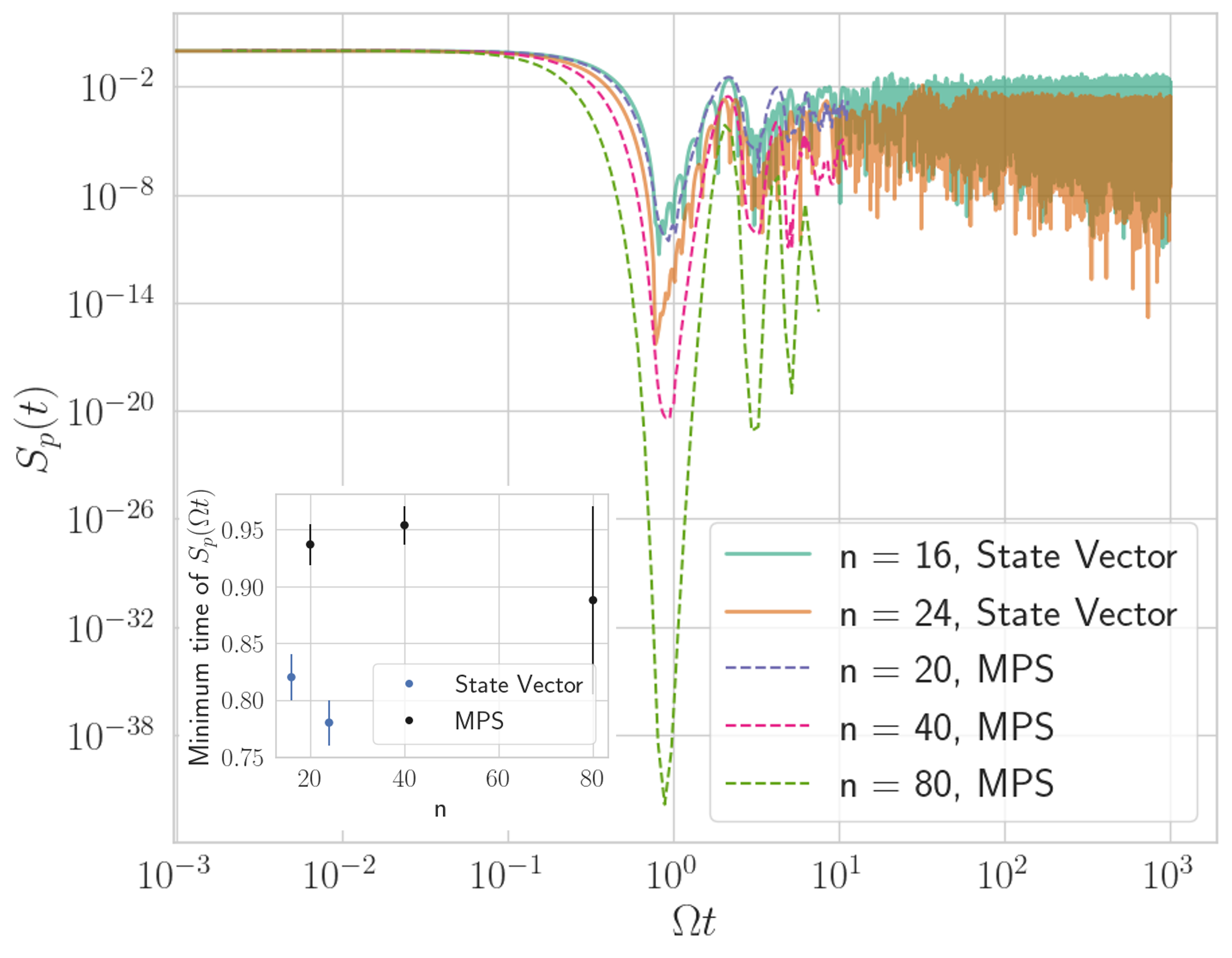}
    \caption{The survival probability \(S_p(t) = |\bra{0}^{\otimes n} \mathcal{U}(t)\ket{0}^{\otimes n}|^2\) of the all-zero state, with unitary evolution \(\mathcal{U}(t) = \exp(-itH_{\mathrm{Ryd}})\), for chains of different lengths up to \(n=80\) qubits. Two types of simulations were performed: a state-vector simulation and an MPS simulation using emu-mps~\cite{bidzhiev2025_emu_sv,weinberg_quspin}. Both show the minimum value in the survival probability occurs at \(\Omega t \sim 0.9\). The inset shows the extracted minimum time, which remains roughly constant.}
    \label{fig:sp_time}
\end{figure}

\begin{figure}
    \centering
    \includegraphics[width=\linewidth]{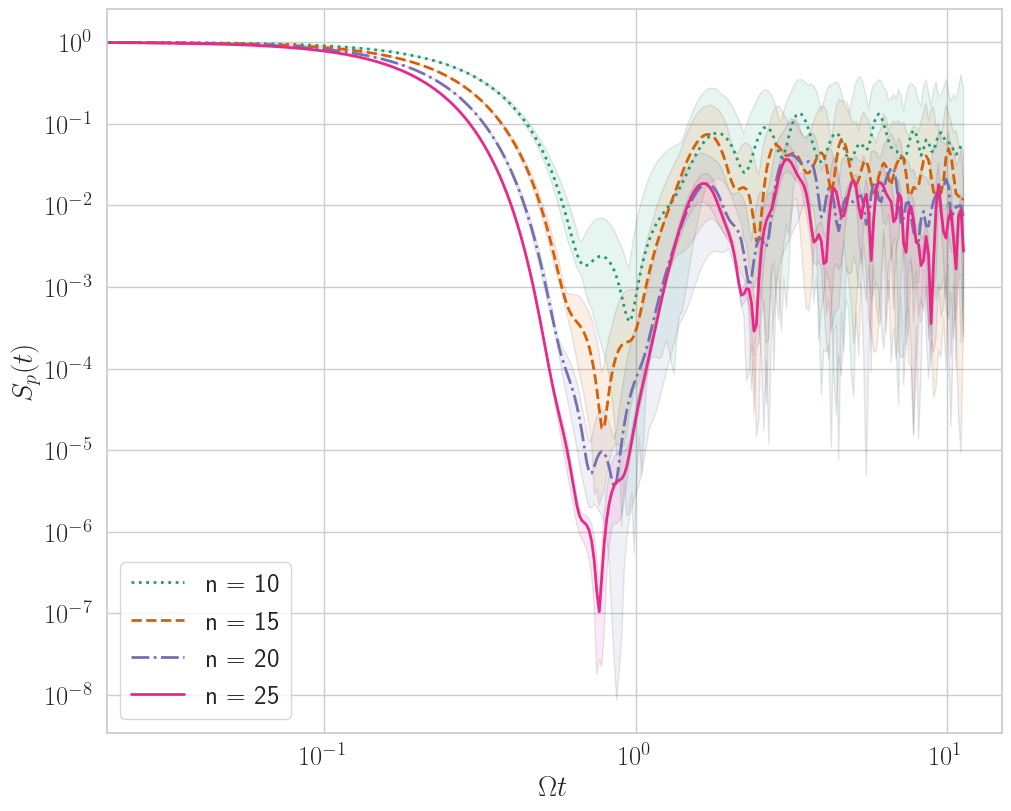}
    \caption{The survival probability \(S_p(t) = |\bra{0}^{\otimes n} \mathcal{U}(t)\ket{0}^{\otimes n}|^2\) of the all-zero state, with unitary evolution \(\mathcal{U}(t) = \exp(-itH_{\mathrm{Ryd}})\), for random punctured triangular grids. For each size, the evolution of the survival probability is averaged over 100 random graphs. Shaded regions show the range of the maximum and minimum values at each point in time for all of the evolutions. This dataset was computed using a GPU-enhanced state-vector emulation tool called emu-sv~\cite{bidzhiev2025_emu_sv}.}
    \label{fig:sp_time_2d}
\end{figure}

The minimum evolution time of our algorithm must be chosen so as to ensure the system evolves long enough for each state to eventually become accessible, while remaining as short as possible. A measure of the relevant time scale is the Thouless time, defined as the time it takes for the survival probability (SP) \(S_p(t) = |\bra{x_i} \mathcal{U}(t)\ket{x_i}|^2\) of an initial state $\ket{x_i}$ evolving under unitary $\mathcal{U}$ to reach a minimum~\cite{LFSantos_thoulesstime}, usually averaged over all $\ket{x_i}$. The Thouless time generally scales exponentially with system size. In contrast, in Figs.~(\ref{fig:initial_state}-\ref{fig:sp_time_2d}) we observe that, under Rydberg-blockaded dynamics, the SP reaches a size- and geometry-independent minimum. We cannot conclude whether the location of the minimum in these figures is the Thouless time. Instead, we use the location of this minimum as an indicator that the initial state amplitude has spread into the ground space of $H_{\mathrm{Ryd}}$. 

First, different initial states are expected to decay at different rates~\cite{schnee2024unconventional}. Under the Rydberg Hamiltonian, the decay will depend on how constrained the initial state is by the Rydberg blockade. For very early times, the SP displays a universal quadratic decay~\cite{Flambaum2001_earlyDecay} with a rate that can be determined by the local density of states (LDOS). This is related to the number of states directly coupled to the initial state~\cite{LFSantos_thoulesstime}.
For a strongly blockaded system, the \(\ket{0}^{\otimes n}\) state is coupled directly to \(n\) other states, but for a state that represents a maximum independent set (IS) this can vary from \(n/2\) for a 1D chain to \(n/3\) for a regular triangular lattice.
For very short times, we can expect that the \(\ket{0}^{\otimes n}\) state will decay the fastest, while a state represented by a maximal IS will decay the slowest. In Fig.~\ref{fig:initial_state}, we see that by \(\Omega t = 10\), all initial states have decayed significantly from their initial configuration for both the chain and the triangular grid. 

Second, this decay feature is expected to be independent of system size, as it correlates with features in the LDOS that are size-independent \cite{schnee2024unconventional}. In Fig.~\ref{fig:sp_time}, for the \(\ket{0}^{\otimes n}\) state on 1D systems of different sizes, we see the minimum value of SP occurs at roughly the same time.

Finally, to ensure that we provide sufficient evolution time in the case of punctured grids, we evolve the \(\ket{0}^{\otimes n}\) state on 100 random punctured triangular graphs with different system sizes and varying connectivity. In Fig.~\ref{fig:sp_time_2d} we again observe that the minimum value of SP is occurring at roughly the same time for all system sizes tested.

While the time at which the SP reaches a minimum is largely independent of system size and graph geometry, it can vary depending on the initial state. To accommodate the slower decay of IS states and give additional time for the SP to evolve past the minimum, we set the minimum evolution time to \(t_{\mathrm{min}} \approx \frac{10}{\Omega}\). Thus, on average, after $t = \frac{10}{\Omega}$, the evolved state amplitudes have spread away from the initial state for all system sizes, geometries, and initial states.

\section{Hardness of punctured grid instances}
\label{appendix:classical}

\begin{figure}
    \centering
    \includegraphics[width=\linewidth]{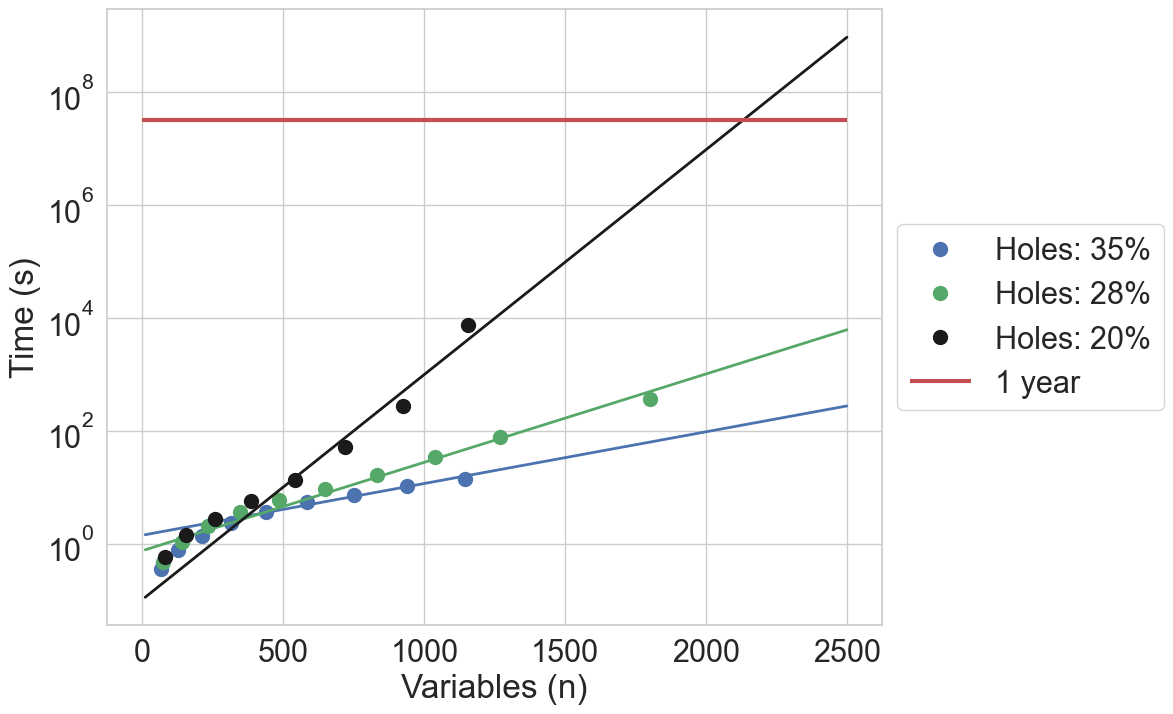}
    \caption{Median runtime of a classical approximate counter (GANAK) for punctured triangular graphs of varying number of variables $n$ and density of holes. There are 1256 instances per data point. For graphs with many holes, the scaling of the median runtime appears sub-exponential. When the density of holes is low, $\leq 20\%$ holes, the runtime scaling appears exponential. Lines are fits to an exponential function using the data collected for the four largest system sizes.}
    \label{fig:classical}
\end{figure}

To estimate the hardness of the punctured triangular grid instances studied in this work, we performed benchmarks using the state-of-the-art classical counter GANAK~\cite{ganak2019}, varying both the number of variables $n$ and the density of holes. Fig.~\ref{fig:classical} shows the median time to compute the count for 1256 random instances per system size, with each color corresponding to a different density of holes ($20\%$, $28\%$ and $35\%$).
For instances with $>20\%$ of sites removed, the scaling appears sub-exponential. Conversely, for instances with $\leq20\%$ of sites removed, the median runtime scales roughly exponentially. This indicates that punctured triangular grid instances are hard on average.

\section{Experimental Protocol}
\label{appendix:exp}
Two independent \textsc{RydCount} experiments were performed using the Fresnel Canada 1 device, an Orion Alpha-class quantum computer manufactured by Pasqal and located in Sherbrooke, Quebec. Both experiments tackled the same 2SAT instance defined on a triangular lattice with $n=81$ sites (see Fig.~\ref{fig:hardware_prob}), whose count of independent sets, computed with the exact SAT counter GANAK~\cite{ganak2019}, is $\sim 4.0879 \times 10^{12}$ (shown in Table~\ref{tab:results}). The algorithm is performed on the QPU as follows:

\begin{itemize}
    \item We preemptively select a separator path on the graph along which the self-reduction steps are performed. Once all self-reduction steps have been completed, the graph is split in two, and a classical counter is used on these smaller and therefore easier graphs. Fixing the reduction path in \textsc{RydCount} does not alter its ability to return a count within the same theoretical bounds.
    \item Two self-reduction paths were tested. The first path is defined by choosing $x_c = 1$ for every variable in the path, which removes both the $c$-th atom and its nearest neighbors from the register. Thus, $5$ steps are required until the instance is fully split in half and we can complete the count using a classical counter. The second path involves choosing $x_c = 0$ instead, which removes only one atom at a time. This leads to a total of $10$ JVV steps along the separator.
    \item Quenches follow the \textsc{RydCount} protocol with $k = 1$. We perform $40$ quenches, whose individual durations $t_{\mathrm{rand}}$ are randomly selected through $t_{\mathrm{rand}} \xleftarrow{\text{sample}} U(t_{\mathrm{min}}=2\,\mu\mathrm{s}, t_{\mathrm{max}} = 6\,\mu\mathrm{s})$ where $U$ is the uniform distribution. For each quench, $200$ measurements are taken in the computational basis. Experimental parameters of $\Omega$ and interatomic spacing are set to $4 \,\text{MHz}$ and $5 \, \mu\text{m}$. This quench is repeated at each self-reduction step. We obtain an ensemble of bitstrings $\sigma$ at each step.
    \item Based on this $\sigma$, we obtain the estimate of the likelihood $p_i \gets \frac{1}{n_{\mathrm{samp}}} \sum_{x\in \sigma} x_i $, where $n_{\mathrm{samp}} = |\sigma|$, for the next node along the separator. These local likelihoods are then used to extract an estimate of the count according to \textsc{RydCount}.
\end{itemize}

Altogether, these results are presented in Fig.~\ref{fig:hardware_count}. We show the relative solution count $C_{\rm est}/C_{\rm exact}$ between the estimated count $C_{\rm est}$ and the exact count $C_{\rm exact}$, as a function of the number of samples taken. This is akin to performing a bootstrapping analysis on the overall samples collected at each step~\cite{Efron1979_bootstraps}.

The complete data is presented in Table~\ref{tab:results}. Notably, the count using the JVV decimation path where we select on $x_c =0$ leads to the closest results to the exact count. This is likely due to the fact that, with this path, we give more of the counting task to the exact  classical counter that completes the count once the instance is separated. Indeed, choosing $x_c = 1$ removes more atoms, and thus the leftover halves each contain fewer variables, while selecting on $x_c =0$ leaves larger sub-instances whose count is obtained classically.

\begin{table}
\centering
\begin{tabular}{|l|c|c|}
\hline
\textbf{Method} & \textbf{Count estimate} & \textbf{Count estimate} \\ 
 & Reduce on $x_c = 1$ & Reduce on $x_c = 0$ \\ \hline
\textsc{RydCount}&&\\ QPU Total Shots & $2.1356 \times 10^{13}$ & $4.0167\times 10^{12}$ \\ \hline
\textsc{RydCount}
 &&\\QPU Postselected Shots & $3.7839 \times 10^{13}$ & $3.2726\times10^{12}$\\ \hline \hline
GANAK (exact) & $4.0879 \times 10^{12}$  & $4.0879 \times 10^{12}$  \\ \hline

\end{tabular}
\caption{Comparison of counting methods for the triangular grid instance of size $n=81$. 
Postselection refers to a new $\sigma'$ where only valid independent sets are kept. 
}
\label{tab:results}
\end{table}

\section{RydCount performance in the limit of perfect Rydberg blockade}
\label{appendix:PXP}

If $V_{ij}$ is taken to be infinitely larger than $\Omega$ and the next nearest neighbors interactions are suppressed, the Rydberg Hamiltonian is approximated by the PXP Hamiltonian~\cite{Lesanovsky2011, Lesanovsky2011a}
\begin{equation}
\label{eq:pxp}
    H_{PXP}(G, \Omega) = \Omega \sum_{i\in V} X_i \prod_{(i,j)\in E} P_{j} \,, 
\end{equation}
with $P_j = \frac{1 - \sigma^z_j}{2} = \mathbb{I} - \ket{1}_j \bra{1}_j$.
The PXP Hamiltonian captures the low-energy dynamics in the Rydberg blockade, where evolution is restricted within the common $+1$ eigenspace of all projectors, namely, the states $\{ \ket{0}_i \ket{0}_j, \ket{0}_i \ket{1}_j, \ket{1}_i \ket{0}_j \}$ for $(i,j) \in E$.
The state $\ket{0}^{\otimes n}$ with all atoms relaxed to their ground state, in which NAQCs are typically initialized, is a $+1$ eigenstate of all projectors. This eigenspace is exactly the solution set of  Eq.~\eqref{eq:satcnf}.

\begin{figure}[t]
    \centering
    \includegraphics[width=\linewidth]{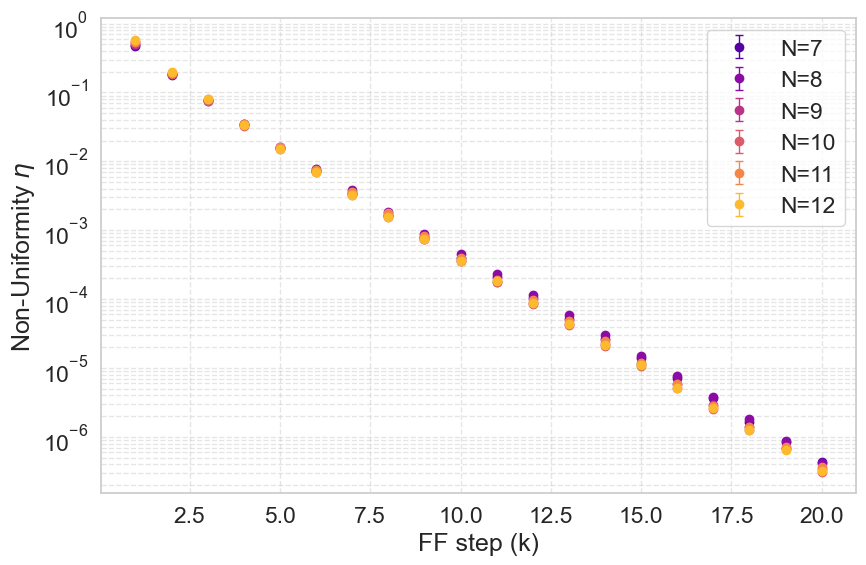}
    \caption{Nonuniformity $\eta$ as a function of number of FF steps for \textsc{RydSamp} applied to a 1D chain instance, using the PXP Hamiltonian.}
    \label{fig:tvd_vs_step_PXP}
\end{figure}

This idealized model provides insight into the performance of the algorithm in a hypothetical scenario where the long-range tail of the dipole-dipole interaction is suppressed.
Fig.~\ref{fig:tvd_vs_step_PXP} shows an exponential decay in the non-uniformity of \textsc{RydSamp} with the PXP Hamiltonian on a 1D chain as a function of the number of FF steps. This suggests that in a perfect Rydberg blockade $n\eta = \mathcal{O}(1)$ is achievable for all $n$.

\begin{figure}[t]
    \centering
    \includegraphics[width=\linewidth]{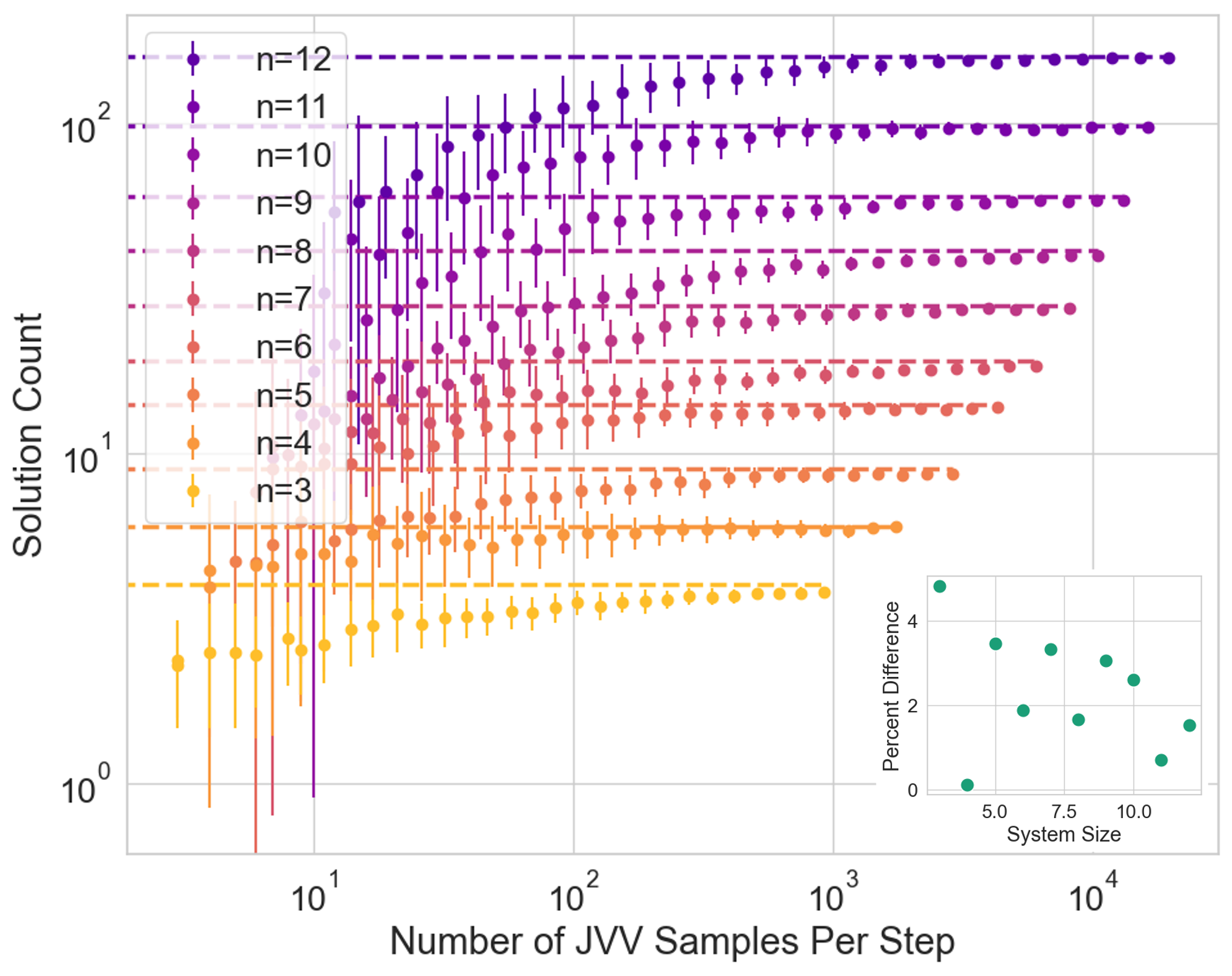}
    \caption{Accuracy of \textsc{RydCount} as a function of $n_{\text{samp}}$ with 15 FF steps evolving under the PXP Hamiltonian for 2D triangular grids of size $n$. Dashed lines represent exact counts. Inset shows the relative error between estimated and exact counts versus system size.}
    \label{fig:compound_PXP}
\end{figure}

\begin{figure}[b]
    \centering
    \includegraphics[width=0.9\linewidth]{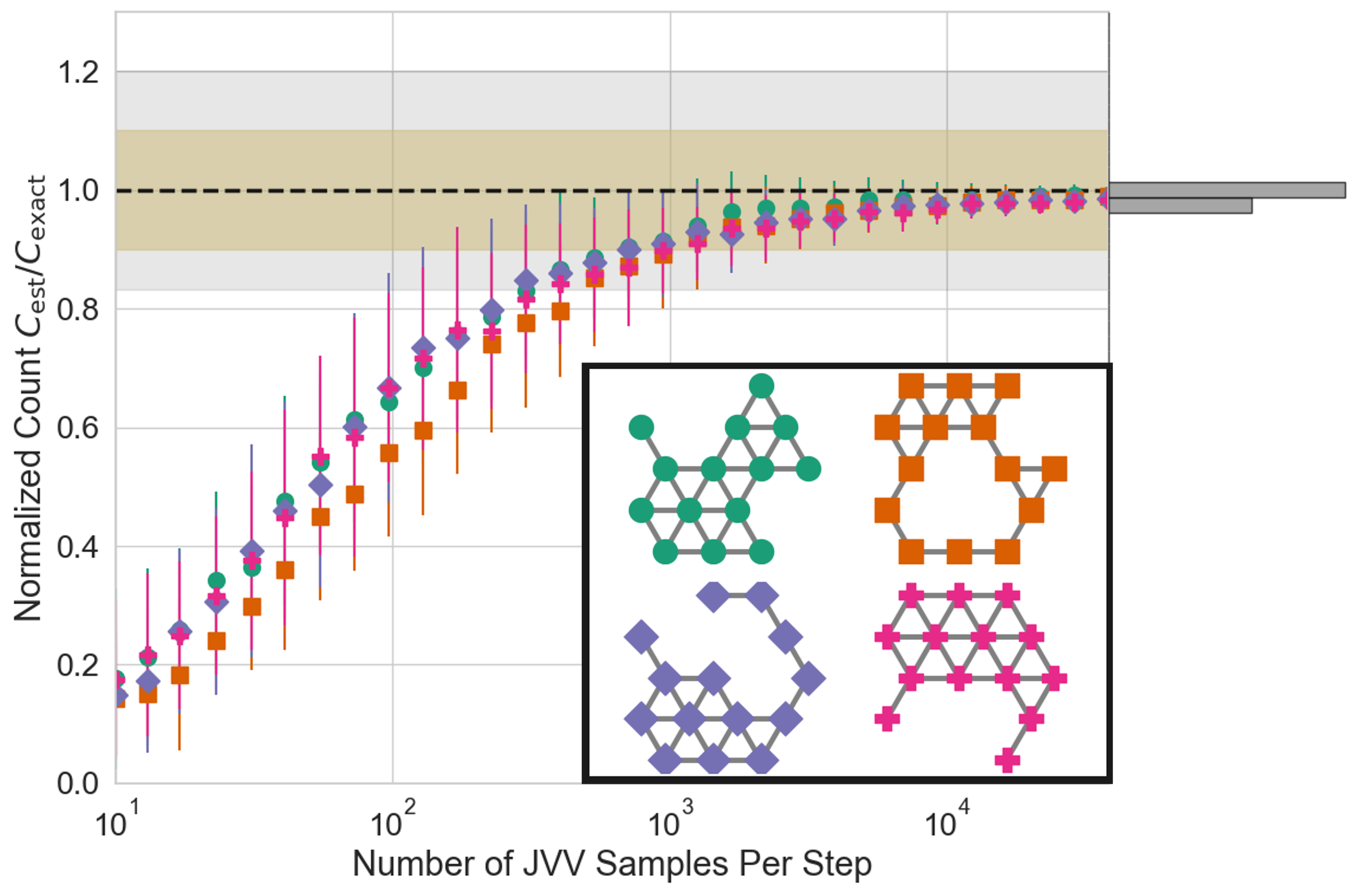}
    \caption{Ratio between the \textsc{RydCount} estimate $C_{\text{est}}$, obtained using the PXP Hamiltonian, and the exact count $C_{\rm exact}$ versus number of samples per self-reduction step for the 4 random punctured triangular grid instances shown in the inset with matching colors and markers. Shaded areas surrounding $C_{\text{est}}/C_{\text{exact}}=1$ correspond to approximate counts with a multiplicative factor of $1.1$ (light yellow) and $1.2$ (light grey). To obtain the histogram, we generate random instances where $4$ vertices are uniformly randomly chosen and removed from a $n=18$ grid. \textsc{RydCount} is performed on 100 of these random instances.}
    \label{fig:2d_jvv_results_punched_PXP}
\end{figure}

This intuition holds beyond 1D. We re-simulate \textsc{RydCount} on 2D triangular grids with the PXP Hamiltonian. Fig.~\ref{fig:compound_PXP} shows that \textsc{RydCount} again accurately estimates total counts. Finally, Fig.~\ref{fig:2d_jvv_results_punched_PXP} shows the convergence of the \textsc{RydCount} algorithm using the PXP Hamiltonian towards the exact count for triangular punctured grids with low hole densities. With the exact same instances as in Fig.~\ref{fig:2d_jvv_results_punched}, the systematic overestimation of $C_{\text{exact}}$ is no longer observed and the distribution of final counts is very sharp around the exact count. Thus, \textsc{RydCount} in the idealized limit converges to $C_\text{exact}$ as predicted by the JVV theorem.

\bibliography{biblio}
\end{document}